\pdfoutput=1

\documentclass[11pt]{article}

\usepackage[preprint]{acl}

\usepackage{times}
\usepackage{latexsym}

\usepackage[T1]{fontenc}

\usepackage[utf8]{inputenc}

\usepackage{microtype}

\usepackage{inconsolata}

\usepackage{graphicx}
\graphicspath{{./Images/}}
\usepackage{hyperref}       %
\usepackage{url}            %
\usepackage{booktabs}       %
\usepackage{amsfonts}       %
\usepackage{nicefrac}       %
\usepackage{microtype}      %
\usepackage{xcolor}         %
\usepackage{colortbl}
\usepackage{comment}
\usepackage{multirow}
\usepackage{adjustbox}
\usepackage{subcaption}
\usepackage{amssymb}
\usepackage{diagbox}

\usepackage{textcomp}

\title{Audio-Aware Large Language Models as Judges for Speaking Styles}

\author{
  \textbf{Cheng-Han Chiang\textsuperscript{1,2}}\thanks{Work done during an internship at Microsoft GenAI. dcml0714@gmail.com},
  \textbf{Xiaofei Wang\textsuperscript{2}}\thanks{Correspondence: xiaofei.wang@microsoft.com},
  \textbf{Chung-Ching Lin\textsuperscript{2}},
  \textbf{Kevin Lin\textsuperscript{2}},\\
  \textbf{Linjie Li\textsuperscript{2}},
  \textbf{Radu Kopetz\textsuperscript{2}},
  \textbf{Yao Qian\textsuperscript{2}},
  \textbf{Zhendong Wang\textsuperscript{2}},\\
  \textbf{Zhengyuan Yang\textsuperscript{2}},
  \textbf{Hung-yi Lee\textsuperscript{1}},
  \textbf{Lijuan Wang\textsuperscript{2}}\\
  \\
  \textsuperscript{1}National Taiwan University,
  \textsuperscript{2}Microsoft, USA\\
}

\begin{document}
\maketitle
\begin{abstract}
Audio-aware large language models (ALLMs) can understand the textual and non-textual information in the audio input.
In this paper, we explore using ALLMs as an automatic judge to assess the speaking styles of speeches.
We use ALLM judges to evaluate the speeches generated by SLMs on two tasks: voice style instruction following and role-playing.
The speaking style we consider includes emotion, volume, speaking pace, word emphasis, pitch control, and non-verbal elements.
We use four spoken language models (SLMs) to complete the two tasks and use humans and ALLMs to judge the SLMs' responses.
We compare two ALLM judges, GPT-4o-audio and Gemini-2.5-pro, with human evaluation results and show that the agreement between Gemini and human judges is comparable to the agreement between human evaluators.
These promising results show that ALLMs can be used as a judge to evaluate SLMs.
Our results also reveal that current SLMs, even GPT-4o-audio, still have room for improvement in controlling the speaking style and generating natural dialogues.
\end{abstract}

\section{Introduction}
\label{section: Introduction}

Ever since the introduction of GPT-4o voice mode~\citep{gpt4o}, the speech community has been moving forward rapidly to propose a model or pipeline that can understand input speech and generate fluent output speech like GPT-4o~\citep{slm_survey}.
The research on this topic can be largely divided into two types: 
(1) \textbf{audio-aware large language models (ALLMs)} that can take texts and audios as input and generate texts~\citep{ltuas,qwen2audio,salmonn}, and 
(2) \textbf{spoken language models (SLMs)} that can take audio and texts as input and generate speech output~\citep{moshi,glm4voice,miniomni,kimiaudio}.
Only recently, some open-source SLMs have shown performance close to GPT-4o~\citep{qwen25omni}.

When building better SLMs, a critical question is how to evaluate the speech they generate.
The textual contents in speech can be easily evaluated by first transcribing the speech into text and evaluating the text using evaluation metrics for text.
For example, using accuracy for question answering (QA)~\citep{voxeval} or using LLM-as-a-judge~\citep{chiang-lee-2023-large} for reference-free evaluation.
Evaluating paralinguistic aspects of the speech, including emotion, prosody, and emphasis, is much more challenging and typically relies on human evaluation~\citep{s2sarena,stepaudio}, which is costly and has been shown to have high variance~\citep{chiang2023we}.

In this paper, we explore whether ALLMs can be an automatic judge of the speaking styles.
To this end, we construct two tasks that test SLM's ability to generate speeches with appropriate speaking styles and evaluate the generated speeches with ALLM judges.
In \textbf{voice style instruction following (voice style IF)}, we instruct SLMs to say a given sentence with fine-grained speaking style instructions.
In \textbf{role-playing}, we prompt an SLM with some role-playing contexts and make it engage in a multi-turn role-playing dialogue.
We evaluate the speech generated by the four SLMs, GPT-4o-audio (4o-audio)~\citep{gpt4o}, GPT-4o-mini-audio (4o-mini-audio), Step-Audio~\citep{stepaudio}, and Qwen-2.5-Omni~\citep{qwen25omni}, using two ALLM judges: GPT-4o-audio~\citep{gpt4o} and Gemini-2.5-Pro~\citep{gemini2.5}.
We compare the evaluation results given by ALLM judges with human evaluation results and show that the Gemini-human judge agreement can be comparable to the human-human judges' agreement.
We summarize our contributions as follows:
\begin{enumerate}
    \item We show that ALLMs can be used as an automatic judge for speaking styles.
    \item We release two tasks for evaluating the speaking style controllability of SLMs, both can be evaluated automatically.
    \item Our results show that SLMs still have room for improvement in speaking style control.
\end{enumerate}

\section{Related Works}
Using large language models (LLMs) as a judge to evaluate the quality of texts, i.e., \textit{LLM-as-a-judge}~\citep{chiang-lee-2023-large,mtbench}, is a mainstream way to evaluate texts in NLP~\citep{alpaca_eval,lc_alpaca}.
However, there has been no success in applying SLMs or ALLMs to evaluate speeches beyond their textual contents. 
\citet{s2sarena} try to use Qwen-2-Audio and 4o-audio as a judge, but find that the results do not align with human evaluation.
\citet{qwenmos} fine-tunes an ALLM on mean opinion score (MOS) datasets; however, the fine-tuned model can only be used for MOS prediction.
We show that ALLMs can be used for more general evaluation beyond MOS prediction, including the style adherence to the instructions and realism of speaking styles.

\section{Task Introduction: StyleSet}
\label{section: BT-Bench}

Our goal is to understand whether ALLMs can evaluate the speaking styles of the speech from SLMs.
Therefore, we construct two tasks that require SLMs to generate speech with proper styles and evaluate the speech using ALLM judges.
We call the two tasks the \underline{StyleSet}. 
Precisely, we use \textbf{voice style instruction following} to evaluate whether SLMs can control the speaking styles when explicitly asked to do so, and we use \textbf{role-playing} to see if SLMs can understand the role-playing contexts and show appropriate speaking styles.
Examples of StyleSet are in Appendix~\ref{subsection: Examples of the Dataset} and Appendix Fig.~\ref{fig:dataset_illustration.pdf}.

\subsection{Voice Style Instruction Following}
\label{subsection: Voice Style Instruction Following}

Voice style IF tests whether SLMs can generate speech that meets the fine-grained speaking style instructions they are given.
This task draws inspiration from instruction following TTS~\citep{instructtts} and several recent SLM benchmarks with voice style IF~\citep{stepaudio,s2sarena}.
The fine-grained speaking styles included in this task are common in realistic dialogues.
If an SLM cannot even speak in a specific speaking style when instructed to do so, we cannot expect it to use that speaking style in a more realistic setting.

We construct 20 diverse instances, each with a sentence to be said and its speaking style.
Unlike the speaking styles used in prior speaking style transfer~\citep{instructvc} or instruction-following TTS~\citep{instructtts,prompttts,cosyvoice}, which mostly give utterance-level instructions on the speaking style, our style instructions cover different granularities and broad aspects of the speaking style.
Some style instructions include changing the volume, pace, or pitch within an utterance, stressing specific words, speaking in a particular emotion, speaking with whipping, sobbing, or stuttering, or inserting non-verbal elements (laughter, sighs, or pauses).
We explain how we prompt SLMs to perform voice style IF in Appendix~\ref{appendix: subsection: dataset construction: Voice Style Instruction Following}.

\textbf{Evaluation}\quad
We evaluate the SLM's output speech using a 5-point Likert scale.
A score of \underline{1} indicates that the speech does not follow the given text. 
Scores from \underline{2} to \underline{5} indicate that the speech follows the text and reflects none, a few, most, or all of the required speaking style elements, respectively.
We give the ALLM judge the text that should be spoken, the desired speaking style, and the generated spoken response, and ask the judge to rate the speech based on the scoring rubrics.
The full ALLM judge prompts are given in Appendix~\ref{appendix section: Evaluation Prompts and Instructions}.

\subsection{Role-Playing}
Role-playing also aims to evaluate whether SLMs can generate speech with appropriate speaking styles.
Unlike voice style IF, which explicitly gives a style instruction, role-playing only gives the SLM a role-playing background and expects the SLMs to generate dialogues that fit in the role with an appropriate speaking style.
The dialogue contexts include the background information of the two roles and the first line in the dialogue.
We create 20 contexts from IEMOCAP~\citep{iemocap}.

\begin{table*}[ht]
    \centering
    
    \begin{adjustbox}{max width=\textwidth}
    \begin{tabular}{l ccc|ccc ccc}
        \toprule
        \multirow{2}{*}{\diagbox[width=5em]{SLM}{Judge}} 
        & \multicolumn{3}{c|}{\underline{\textbf{Voice Style IF}}}
        & \multicolumn{6}{c}{\underline{\textbf{Role-Playing}}} \\
        & 4o-audio   & Gemini & Human
        & 4o-audio   & Gemini & Human
        & 4o-audio   & Gemini & Human \\
        \midrule
        Metrics
        & \multicolumn{3}{c|}{\textit{Style: Likert (1-5)} }
        & \multicolumn{3}{c}{\textit{Style: Likert: (1-5)}}
        & \multicolumn{3}{c}{\textit{Realism: (0 / 1)}} \\
        \midrule
        Human
        & --   & --     & --
        & 4.32\textsubscript{0.43} & 4.65\textsubscript{0.56}& 4.03\textsubscript{0.86}
        &      0.97\textsubscript{0.07} & 0.99\textsubscript{0.04} & 0.95\textsubscript{0.10} \\
        4o-audio
        & 3.71\textsubscript{1.19} & 3.83\textsubscript{1.29} & 3.65\textsubscript{1.51}
        & 4.04\textsubscript{0.68} & 4.34\textsubscript{1.16}& 3.39\textsubscript{0.55}
        &      0.80\textsubscript{0.27}& 0.60\textsubscript{0.35} & 0.51\textsubscript{0.15} \\
        4o-mini-audio
        & 2.35\textsubscript{1.28} & 2.29\textsubscript{1.68} & 2.39\textsubscript{1.51}
        & 4.18\textsubscript{0.40} & 4.13\textsubscript{1.47}& 3.29\textsubscript{0.74}
        &      0.96\textsubscript{0.08} & 0.64\textsubscript{0.34} & 0.43\textsubscript{0.22} \\
        Step-Audio
        & 2.71\textsubscript{1.20} & 2.96\textsubscript{1.52} & 2.30\textsubscript{1.20}
        &      3.87\textsubscript{0.40} &                          3.26\textsubscript{1.10}
        & 2.28\textsubscript{0.44}
        &      0.73\textsubscript{0.30}& 0.01\textsubscript{0.04} & 0.25\textsubscript{0.14} \\
        Qwen
        & 2.88\textsubscript{1.53} & 2.46\textsubscript{1.51} & 2.15\textsubscript{1.31}
        &      3.62\textsubscript{0.73} &                          3.09\textsubscript{1.54} & 2.40\textsubscript{0.73}
        &      0.63\textsubscript{0.39}& 0.05\textsubscript{0.11} & 0.29\textsubscript{0.23} \\
        \bottomrule
    \end{tabular}
    \end{adjustbox}
    \caption{Average score and standard deviation (in subscript) per SLM rated by three types of judges.}
    \label{tab:vsif-role}
\end{table*}

Given an SLM, we simulate a dialogue with two roles using that SLM.
The SLM will switch between two roles to speak and form a multi-turn dialogue.
The details on how to prompt a single SLM to complete a role-play with two roles are presented in Appendix~\ref{subsubsection: Prompting an SLM to Form a  Dialogue}.
We concatenate the speeches generated by the two roles to form a dialogue and crop a one-minute audio for evaluation.
While role-playing is not a new task to evaluate SLMs~\citep{stepaudio,s2sarena}, using an SLM to role-play with itself and evaluate the resulting dialogue is not seen in past literature.

\textbf{Evaluation}\quad
We evaluate the spoken dialogue based on two aspects: (1) \textbf{style} and (2) \textbf{realism}.
The reason to separately evaluate these two aspects is that even if the styles sound natural, the whole dialogue may still not sound realistic, as we show in Section~\ref{subsubsection: Results: Role-Playing}.
The \textbf{style} aspect evaluates whether the SLM generates dialogues whose speaking styles fit in the context.
We give the ALLM judge the one-minute audio, the role-playing context, and ask it to evaluate based on the following 5-point Likert scale:
A score of 1 means the SLM fails to complete the role-playing task.
Scores from 2 to 5 indicate that the SLM stays in character and the semantic content is appropriate, while the speaking style ranges from poor (\underline{2}) to very natural (\underline{5}).

For \textbf{realism}, we evaluate whether the role-playing dialogue is like a realistic dialogue between two humans.
As \textit{'realism'} lacks well-defined intermediate levels, we use a binary judgment to avoid ambiguous scoring:
A score of \underline{0} indicates the dialogue is unlikely to be human-generated, while \underline{1} suggests it is likely to be human-generated.
The full judge prompts are in Appendix~\ref{appendix section: Evaluation Prompts and Instructions}.

\section{Experiment}
\label{section: Experiment}

\subsection{Experiment Setup}
\label{subsection: Experiment Setup}
We use four SLMs to complete the tasks in StyleSet and use two ALLM judges for evaluation.
We select 4o-audio and Gemini-2.5-pro~\citep{gemini2.5} as the ALLM judges\footnote{We refer to them as ALLM judges as we only generate texts from them, while they can output speeches.}.
The SLMs that are used to complete the tasks in StyleSet include 4o-audio, 4o-mini-audio~\citep{gpt4o}, Step-Audio~\citep{stepaudio}, and Qwen-2.5-Omni~\citep{qwen25omni}.
The selection of these models is because they are publicly available and support multi-turn dialogue.

When the ALLM judge generates the evaluation output, we allow it to generate chain-of-thought (CoT) reasoning~\citep{cot}, which has been shown to increase the agreement between LLM and human judges~\citep{closer}.
We use regular expressions to extract the numeric prediction from the judge's output.
For each instance to be evaluated, we sample five judge responses and ensemble the verdicts~\citep{selfconsistency}.
The hyperparameters we use are in Appendix~\ref{section: Experiment Details}.

To justify using ALLMs as judges, we compare the results of ALLM judges with human evaluation results.
We recruit four human evaluators for each task and ask them to run the same evaluation pipeline as what ALLM judges do.
We keep the instructions to human evaluators and ALLM judges as similar as possible.
The details of the human evaluation are in Appendix~\ref{subsection: Details of Human Evaluation}.

\subsection{Results: Voice Style IF}
\label{subsubsection: Results: Voice Instruction}
Table~\ref{tab:vsif-role} shows the average score per model given by three different types of judges: 4o, Gemini, and humans.
We have the following observations:

\textbf{Human rates 4o-audio the highest and the other three comparably low.}\quad
We focus on the human evaluation results first, as this can be considered the \textit{ground truth} for the evaluation results.
4o-audio is the best, with an average score of 3.65, indicating that it can follow some or most of the style instructions, but is still far from perfect.
The performance of the other three SLMs is rather close and bad, with Step-Audio closely following 4o-mini and Qwen the worst.
By analyzing the instances in which the SLMs perform badly, we find that all models cannot vary the speaking pace within an utterance.
We also see Qwen-2.5-Omni and Step-Audio sometimes cannot insert non-verbal elements like laughter or sighs, but directly reads the words "sigh" out loud.

\textbf{ALLM judges also rate 4o-audio the highest.}\quad
We shift our focus to the automatic evaluation given by the two ALLM judges.
Similar to the result of human evaluation, 4o-audio is rated as the best model by the two ALLM judges.
While using 4o-audio to judge itself may be prone to self-enhancement bias~\citep{mtbench}, the results from human evaluation show that 4o is indeed the best, and Gemini-2.5-pro also agrees with this verdict.
The ranking of the remaining three SLMs is more inconsistent between the ALLM judges and human evaluators.
However, this is not surprising since the average scores of the three models are very close in human evaluation.
These results indicate that while ALLM judges can distinguish good SLMs from the bad ones, it may be hard to compare several bad ones.

\begin{table}[t!]
    \centering
    \begin{adjustbox}{max width=\columnwidth}
        \begin{tabular}{l c c }
            \toprule
            & Voice Style IF        & Role-Playing (Style)       \\
            \midrule
            \textit{Human–Human}   & 0.596            &                    0.253\\
            \textit{Human–4o}      &                  0.355&                    0.305\\
            \textit{Human–Gemini}  & 0.640            &                    0.319\\
            \bottomrule
        \end{tabular}
    \end{adjustbox}
    \caption{The average Pearson's $r$ between two judges.}
    \label{tab:agreement}
\end{table}

\textbf{Score correlation between Gemini and human judges is high}.\quad
An alternative and commonly used way to evaluate the agreement of judges is the correlation coefficient between the scores given by two judges~\citep{amidei-etal-2019-agreement}.
The correlation coefficient between two evaluators is computed over two arrays with 80 scores (4 models $\times$ 20 instances).
In Table~\ref{tab:agreement}, we report the Pearson's $r$ of the scores between a pair of human evaluators, and the average correlation coefficient between the ALLM judge and each human evaluator.
The average Pearson's $r$ between human evaluators is 0.596, which is reasonably high, validating the quality of human evaluation.
For the ALLM-human judge correlation, Gemini achieves an average correlation of 0.640 with human evaluators, even higher than the pairwise correlation between humans.
The 4o judge has a much lower Pearson's $r$ with the human evaluators, only 0.355.
This validates that using Gemini as a judge on this task can obtain results that are close to human evaluation.

\textbf{Gemini judge's variance due to hyperparameters is low}.\quad
We evaluate the same sets of SLM outputs with Gemini using three temperatures (1.5, 1.0, 0.5) when generating the judge responses.
We use Gemini here as it has a higher correlation with humans.
The human-Gemini Pearson's $r$'s vary between 0.640 to 0.649, which is rather stable.

\subsection{Results: Role-Playing}
\label{subsubsection: Results: Role-Playing}

The results for role-playing are given in the right portion of Table~\ref{tab:vsif-role}.
We also evaluate the human-recorded dialogues in IEMOCAP using human and ALLM judges; the results are in the top row in Table~\ref{tab:vsif-role}.
We have the following findings:

\paragraph{Humans rate human-recorded dialogue higher than 4o-generated ones.}\quad
Humans rate the human-recorded dialogues with an average rating of 4.03, significantly higher than all the SLMs.
On the 5-point \textit{style} aspect, the average scores of human-recorded dialogues and 4o-generated role-plays only differ by 0.64, which seems rather small under the 5-point scale.
However, the \textit{realism} rating shows that human-recorded dialogues are much realistic, with a realism score almost twice as much as 4o.
This shows that current SLMs are still not good enough to generate realistic dialogues.

\paragraph{Gemini judge generally agrees with human judges.}\quad
Gemini judge also rates human-recorded dialogue as the best style and being the most realistic, and 4o-audio being the best SLM but lagging behind humans.
However, the gap between 4o and 4o-mini is not very significant, agreeing with the human evaluation results.
Qwen-2.5-Omni and Step-Audio are worse than the two 4o-series models; their performance is similar, making it difficult to determine which one is better, which also aligns with the human evaluation results.
For the 4o judge, the exact rankings among SLMs slightly disagree with human results while maintaining the general trend that humans are better than the 4o series models, and open-sourced SLMs are the worst.

\paragraph{Human-SLM judge correlations are reasonable.}
\quad
We consider the Pearson's $r$ of the 5-point \textit{style} aspect.
The average pairwise human-human Pearson's $r$ is only 0.253, showing that evaluating the styles of dialogue can be somewhat subjective, but some weak agreement still exists among human evaluators.
The average human-4o correlation and human-Gemini correlation are higher than 0.30, exceeding the human-human correlation.
This shows that evaluating role-playing with ALLM judges is at least as good as using human evaluators.

\section{Conclusion}
\label{section: Conclusion}
This paper attempts to use ALLM to judge the speaking styles generated by SLMs.
We use two tasks, voice style IF and role-playing, to generate speeches from SLMs that have diverse speaking styles and evaluate those speeches with ALLMs.
By comparing the evaluation results from human and ALLM judges, we find that ALLMs can be used as automatic judges on these two tasks and achieve agreement with human judges comparable to the agreement within human judges. 

\section*{Limitations}
We see the following limitations of this paper.
First, we only use ALLM judges to evaluate speaking styles, while there are many different attributes in speech that can be evaluated.
As a result, the conclusion of this paper cannot and should not be taken as "\textit{ALLM judges can be used to evaluate SLMs}"; we only validate the effectiveness of ALLM judges on assessing speaking styles.
We recommend that future research be done on evaluating other aspects of speech using ALLM judges.

Next, in the evaluation of role-playing, we let the SLM speak turn by turn.
This turn-taking dialogue may be different from the full-duplex dialogue between humans.
We do not consider full-duplex settings since most SLMs are not capable of doing this. 
However, the evaluation pipeline we propose naturally supports evaluating full-duplex dialogue.
It will be interesting to see how making the dialogue full-duplex increases the dialogue's realism for the ALLM judges, and we leave it as a future work when more powerful SLMs can operate in full-duplex.

Thirdly, we only consider single-wise (point-wise) evaluation, which assigns a score to an instance to be evaluated.
We do not study pairwise comparison, which gives the judge two instances to evaluate and ask which one is better.
The reason to focus on single-wise evaluation is that pairwise evaluation may be hard, even for humans, when the two instances for comparison are equally bad.
For example, the role-play dialogues generated by Qwen-2.5-Omni and Step-Audio sound equally bad, and it is hard to select which one is better.

Last, we only evaluate speaking styles in English, while the two SLMs we use, Step-Audio and Qwen-2.5-Omni, may have a stronger ability in Chinese, given that they are trained on massive Chinese data.

We do not see specific harm in our paper.

\section*{Acknowledgment}
We want to thank Yi-Cheng Lin and Yu-Xiang Lin for their valuable insights on our projects.

\bibliography{custom}

\begin{thebibliography}{30}
\providecommand{\natexlab}[1]{#1}

\bibitem[{Amidei et~al.(2019)Amidei, Piwek, and Willis}]{amidei-etal-2019-agreement}
Jacopo Amidei, Paul Piwek, and Alistair Willis. 2019.
\newblock \href {https://doi.org/10.18653/v1/W19-8642} {Agreement is overrated: A plea for correlation to assess human evaluation reliability}.
\newblock In \emph{Proceedings of the 12th International Conference on Natural Language Generation}, pages 344--354, Tokyo, Japan. Association for Computational Linguistics.

\bibitem[{Arora et~al.(2025)Arora, Chang, Chien, Peng, Wu, Adi, Dupoux, Lee, Livescu, and Watanabe}]{slm_survey}
Siddhant Arora, Kai-Wei Chang, Chung-Ming Chien, Yifan Peng, Haibin Wu, Yossi Adi, Emmanuel Dupoux, Hung-Yi Lee, Karen Livescu, and Shinji Watanabe. 2025.
\newblock On the landscape of spoken language models: A comprehensive survey.
\newblock \emph{arXiv preprint arXiv:2504.08528}.

\bibitem[{Busso et~al.(2008)Busso, Bulut, Lee, Kazemzadeh, Mower, Kim, Chang, Lee, and Narayanan}]{iemocap}
Carlos Busso, Murtaza Bulut, Chi-Chun Lee, Abe Kazemzadeh, Emily Mower, Samuel Kim, Jeannette~N Chang, Sungbok Lee, and Shrikanth~S Narayanan. 2008.
\newblock Iemocap: Interactive emotional dyadic motion capture database.
\newblock \emph{Language resources and evaluation}, 42:335--359.

\bibitem[{Chen et~al.(2025)Chen, Hu, Wang, Wang, Chen, Zhang, Yang, and Chng}]{qwenmos}
Chen Chen, Yuchen Hu, Siyin Wang, Helin Wang, Zhehuai Chen, Chao Zhang, Chao-Han~Huck Yang, and EngSiong Chng. 2025.
\newblock \href {https://openreview.net/forum?id=U42TkrEDzb} {Audio large language models can be descriptive speech quality evaluators}.
\newblock In \emph{The Thirteenth International Conference on Learning Representations}.

\bibitem[{Chiang et~al.(2023)Chiang, Huang, and Lee}]{chiang2023we}
Cheng-Han Chiang, Wei-Ping Huang, and Hung-yi Lee. 2023.
\newblock Why we should report the details in subjective evaluation of tts more rigorously.
\newblock In \emph{Proc. Interspeech 2023}, pages 5551--5555.

\bibitem[{Chiang and Lee(2023{\natexlab{a}})}]{chiang-lee-2023-large}
Cheng-Han Chiang and Hung-yi Lee. 2023{\natexlab{a}}.
\newblock \href {https://doi.org/10.18653/v1/2023.acl-long.870} {Can large language models be an alternative to human evaluations?}
\newblock In \emph{Proceedings of the 61st Annual Meeting of the Association for Computational Linguistics (Volume 1: Long Papers)}, pages 15607--15631, Toronto, Canada. Association for Computational Linguistics.

\bibitem[{Chiang and Lee(2023{\natexlab{b}})}]{closer}
Cheng-Han Chiang and Hung-yi Lee. 2023{\natexlab{b}}.
\newblock \href {https://doi.org/10.18653/v1/2023.findings-emnlp.599} {A closer look into using large language models for automatic evaluation}.
\newblock In \emph{Findings of the Association for Computational Linguistics: EMNLP 2023}, pages 8928--8942, Singapore. Association for Computational Linguistics.

\bibitem[{Chu et~al.(2024)Chu, Xu, Yang, Wei, Wei, Guo, Leng, Lv, He, Lin et~al.}]{qwen2audio}
Yunfei Chu, Jin Xu, Qian Yang, Haojie Wei, Xipin Wei, Zhifang Guo, Yichong Leng, Yuanjun Lv, Jinzheng He, Junyang Lin, and 1 others. 2024.
\newblock Qwen2-audio technical report.
\newblock \emph{arXiv preprint arXiv:2407.10759}.

\bibitem[{Cui et~al.(2025)Cui, Jiao, Meng, and King}]{voxeval}
Wenqian Cui, Xiaoqi Jiao, Ziqiao Meng, and Irwin King. 2025.
\newblock Voxeval: Benchmarking the knowledge understanding capabilities of end-to-end spoken language models.
\newblock \emph{arXiv preprint arXiv:2501.04962}.

\bibitem[{D\'efossez et~al.(2024)D\'efossez, Mazar\'e, Orsini, Royer, P\'erez, J\'egou, Grave, and Zeghidour}]{moshi}
Alexandre D\'efossez, Laurent Mazar\'e, Manu Orsini, Am\'elie Royer, Patrick P\'erez, Herv\'e J\'egou, Edouard Grave, and Neil Zeghidour. 2024.
\newblock \href {https://arxiv.org/abs/2410.00037} {Moshi: a speech-text foundation model for real-time dialogue}.
\newblock Technical report, Kyutai Labs.

\bibitem[{Du et~al.(2024)Du, Chen, Zhang, Hu, Lu, Yang, Hu, Zheng, Gu, Ma et~al.}]{cosyvoice}
Zhihao Du, Qian Chen, Shiliang Zhang, Kai Hu, Heng Lu, Yexin Yang, Hangrui Hu, Siqi Zheng, Yue Gu, Ziyang Ma, and 1 others. 2024.
\newblock Cosyvoice: A scalable multilingual zero-shot text-to-speech synthesizer based on supervised semantic tokens.
\newblock \emph{arXiv preprint arXiv:2407.05407}.

\bibitem[{Dubois et~al.(2024)Dubois, Galambosi, Liang, and Hashimoto}]{lc_alpaca}
Yann Dubois, Bal{\'a}zs Galambosi, Percy Liang, and Tatsunori~B Hashimoto. 2024.
\newblock Length-controlled alpacaeval: A simple way to debias automatic evaluators.
\newblock \emph{arXiv preprint arXiv:2404.04475}.

\bibitem[{Gong et~al.(2024)Gong, Luo, Liu, Karlinsky, and Glass}]{ltuas}
Yuan Gong, Hongyin Luo, Alexander~H. Liu, Leonid Karlinsky, and James~R. Glass. 2024.
\newblock \href {https://openreview.net/forum?id=nBZBPXdJlC} {Listen, think, and understand}.
\newblock In \emph{The Twelfth International Conference on Learning Representations}.

\bibitem[{Google(2025)}]{gemini2.5}
Google. 2025.
\newblock \href {https://blog.google/technology/google-deepmind/gemini-model-thinking-updates-march-2025/} {Gemini 2.5: Our most intelligent ai model}.
\newblock Accessed on May 15, 2025.

\bibitem[{Guo et~al.(2023)Guo, Leng, Wu, Zhao, and Tan}]{prompttts}
Zhifang Guo, Yichong Leng, Yihan Wu, Sheng Zhao, and Xu~Tan. 2023.
\newblock Prompttts: Controllable text-to-speech with text descriptions.
\newblock In \emph{ICASSP 2023-2023 IEEE International Conference on Acoustics, Speech and Signal Processing (ICASSP)}, pages 1--5. IEEE.

\bibitem[{Holtzman et~al.(2020)Holtzman, Buys, Du, Forbes, and Choi}]{neucleus}
Ari Holtzman, Jan Buys, Li~Du, Maxwell Forbes, and Yejin Choi. 2020.
\newblock \href {https://openreview.net/forum?id=rygGQyrFvH} {The curious case of neural text degeneration}.
\newblock In \emph{International Conference on Learning Representations}.

\bibitem[{Huang et~al.(2025)Huang, Wu, Wang, Yan, Hu, Feng, Tian, Shen, Li, Chen, Liu, Miao, You, Chen, Yang, Huang, Zhang, Gong, Zhang, Li, Wan, Hu, Ming, Yuan, Zhang, Zhou, Li, Ma, An, Ji, Li, Wen, Ma, Liang, Mou, Ahmidi, Wang, Li, Miao, Xu, Feng, Wang, Shi, Sun, Hu, Sai, Liu, Huang, Yan, Wang, Jia, Zhang, Gong, Wu, Liu, Sun, Zhen, Feng, Wu, Wu, Yang, Wang, Zhang, Lin, Li, Xia, Zhou, Gu, Chen, Wu, Li, Li, Liang, Wang, Hao, Wu, Tan, Pang, Yang, Gao, Liu, Liu, Cao, Wang, Deng, He, Sun, Han, Deng, Liu, Zhao, Wei, Yu, Cao, Li, Ma, Xu, Shi, Wang, Zhong, Luo, Lu, Yin, Yan, Yang, Xie, Ge, Sun, Huang, Chang, Yang, Zhang, Jiao, Jiang, Shum, Chen, Li, Zhou, Zhang, Zhang, and Zhu}]{stepaudio}
Ailin Huang, Boyong Wu, Bruce Wang, Chao Yan, Chen Hu, Chengli Feng, Fei Tian, Feiyu Shen, Jingbei Li, Mingrui Chen, Peng Liu, Ruihang Miao, Wang You, Xi~Chen, Xuerui Yang, Yechang Huang, Yuxiang Zhang, Zheng Gong, Zixin Zhang, and 102 others. 2025.
\newblock \href {https://arxiv.org/abs/2502.11946} {Step-audio: Unified understanding and generation in intelligent speech interaction}.
\newblock \emph{Preprint}, arXiv:2502.11946.

\bibitem[{Jiang et~al.(2025)Jiang, Lin, Bu, Du, Wang, and Li}]{s2sarena}
Feng Jiang, Zhiyu Lin, Fan Bu, Yuhao Du, Benyou Wang, and Haizhou Li. 2025.
\newblock S2s-arena, evaluating speech2speech protocols on instruction following with paralinguistic information.
\newblock \emph{arXiv preprint arXiv:2503.05085}.

\bibitem[{KimiTeam et~al.(2025)KimiTeam, Ding, Ju, Leng, Liu, Liu, Shang, Shen, Song, Tan, Tang, Wang, Wei, Xin, Xu, Yu, Zhang, Zhou, Charles, Chen, Chen, Du, He, Hu, Lai, Li, Liu, Sun, Wang, Wang, Wu, Wu, Yang, Yang, Yang, Yang, Yin, Yuan, Zhang, and Zhou}]{kimiaudio}
KimiTeam, Ding Ding, Zeqian Ju, Yichong Leng, Songxiang Liu, Tong Liu, Zeyu Shang, Kai Shen, Wei Song, Xu~Tan, Heyi Tang, Zhengtao Wang, Chu Wei, Yifei Xin, Xinran Xu, Jianwei Yu, Yutao Zhang, Xinyu Zhou, Y.~Charles, and 21 others. 2025.
\newblock \href {https://arxiv.org/abs/2504.18425} {Kimi-audio technical report}.
\newblock \emph{Preprint}, arXiv:2504.18425.

\bibitem[{Kuan et~al.(2023)Kuan, Li, Hsu, Lin, Chung, Chang, Chang, and Lee}]{instructvc}
Chun-Yi Kuan, Chen-An Li, Tsu-Yuan Hsu, Tse-Yang Lin, Ho-Lam Chung, Kai-Wei Chang, Shuo-Yiin Chang, and Hung-yi Lee. 2023.
\newblock Towards general-purpose text-instruction-guided voice conversion.
\newblock In \emph{2023 IEEE Automatic Speech Recognition and Understanding Workshop (ASRU)}, pages 1--8. IEEE.

\bibitem[{Li et~al.(2023)Li, Zhang, Dubois, Taori, Gulrajani, Guestrin, Liang, and Hashimoto}]{alpaca_eval}
Xuechen Li, Tianyi Zhang, Yann Dubois, Rohan Taori, Ishaan Gulrajani, Carlos Guestrin, Percy Liang, and Tatsunori~B. Hashimoto. 2023.
\newblock Alpacaeval: An automatic evaluator of instruction-following models.
\newblock \url{https://github.com/tatsu-lab/alpaca_eval}.

\bibitem[{OpenAI(2024)}]{gpt4o}
OpenAI. 2024.
\newblock \href {https://openai.com/index/hello-gpt-4o/} {Hello gpt-4o}.
\newblock Accessed on May 12, 2025.

\bibitem[{Tang et~al.(2024)Tang, Yu, Sun, Chen, Tan, Li, Lu, MA, and Zhang}]{salmonn}
Changli Tang, Wenyi Yu, Guangzhi Sun, Xianzhao Chen, Tian Tan, Wei Li, Lu~Lu, Zejun MA, and Chao Zhang. 2024.
\newblock \href {https://openreview.net/forum?id=14rn7HpKVk} {{SALMONN}: Towards generic hearing abilities for large language models}.
\newblock In \emph{The Twelfth International Conference on Learning Representations}.

\bibitem[{Wang et~al.(2023)Wang, Wei, Schuurmans, Le, Chi, Narang, Chowdhery, and Zhou}]{selfconsistency}
Xuezhi Wang, Jason Wei, Dale Schuurmans, Quoc~V Le, Ed~H. Chi, Sharan Narang, Aakanksha Chowdhery, and Denny Zhou. 2023.
\newblock \href {https://openreview.net/forum?id=1PL1NIMMrw} {Self-consistency improves chain of thought reasoning in language models}.
\newblock In \emph{The Eleventh International Conference on Learning Representations}.

\bibitem[{Wei et~al.(2022)Wei, Wang, Schuurmans, Bosma, Xia, Chi, Le, Zhou et~al.}]{cot}
Jason Wei, Xuezhi Wang, Dale Schuurmans, Maarten Bosma, Fei Xia, Ed~Chi, Quoc~V Le, Denny Zhou, and 1 others. 2022.
\newblock Chain-of-thought prompting elicits reasoning in large language models.
\newblock \emph{Advances in neural information processing systems}, 35:24824--24837.

\bibitem[{Xiezhifei(2024)}]{miniomni}
Xiezhifei. 2024.
\newblock \href {https://openreview.net/forum?id=5GwEMXzBOP} {Mini-omni: Language models can hear, talk while thinking in streaming}.
\newblock In \emph{Submitted to Tsinghua University Course: Advanced Machine Learning}.
\newblock Under review.

\bibitem[{Xu et~al.(2025)Xu, Guo, He, Hu, He, Bai, Chen, Wang, Fan, Dang et~al.}]{qwen25omni}
Jin Xu, Zhifang Guo, Jinzheng He, Hangrui Hu, Ting He, Shuai Bai, Keqin Chen, Jialin Wang, Yang Fan, Kai Dang, and 1 others. 2025.
\newblock Qwen2. 5-omni technical report.
\newblock \emph{arXiv preprint arXiv:2503.20215}.

\bibitem[{Yang et~al.(2024)Yang, Liu, Huang, Weng, and Meng}]{instructtts}
Dongchao Yang, Songxiang Liu, Rongjie Huang, Chao Weng, and Helen Meng. 2024.
\newblock Instructtts: Modelling expressive tts in discrete latent space with natural language style prompt.
\newblock \emph{IEEE/ACM Transactions on Audio, Speech, and Language Processing}.

\bibitem[{Zeng et~al.(2024)Zeng, Du, Liu, Wang, Jiang, Zhao, Dong, and Tang}]{glm4voice}
Aohan Zeng, Zhengxiao Du, Mingdao Liu, Kedong Wang, Shengmin Jiang, Lei Zhao, Yuxiao Dong, and Jie Tang. 2024.
\newblock Glm-4-voice: Towards intelligent and human-like end-to-end spoken chatbot.
\newblock \emph{arXiv preprint arXiv:2412.02612}.

\bibitem[{Zheng et~al.(2023)Zheng, Chiang, Sheng, Zhuang, Wu, Zhuang, Lin, Li, Li, Xing, Zhang, Gonzalez, and Stoica}]{mtbench}
Lianmin Zheng, Wei-Lin Chiang, Ying Sheng, Siyuan Zhuang, Zhanghao Wu, Yonghao Zhuang, Zi~Lin, Zhuohan Li, Dacheng Li, Eric Xing, Hao Zhang, Joseph~E. Gonzalez, and Ion Stoica. 2023.
\newblock \href {https://openreview.net/forum?id=uccHPGDlao} {Judging {LLM}-as-a-judge with {MT}-bench and chatbot arena}.
\newblock In \emph{Thirty-seventh Conference on Neural Information Processing Systems Datasets and Benchmarks Track}.

\end{thebibliography}

\appendix

\begin{figure*}[t]

\centering
\includegraphics[clip, trim = 0px 320px 0px 0px,width=0.9\linewidth]{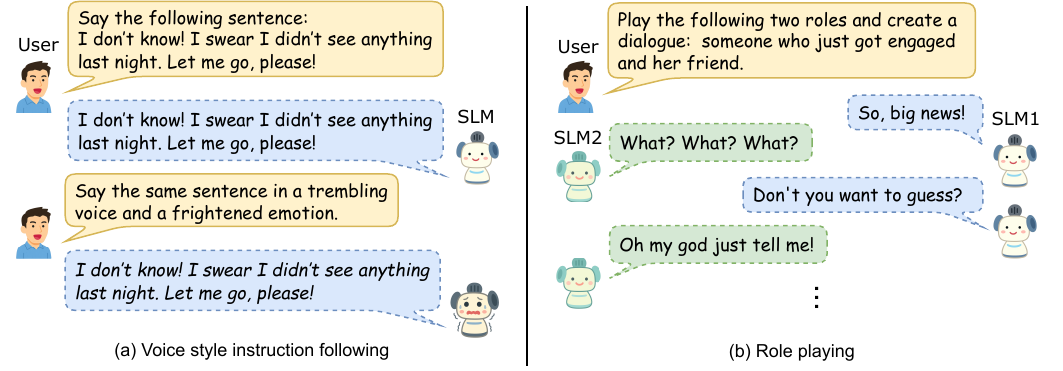}
\caption{An illustration of the dataset. 
(a) Voice style IF: We only evaluate the last turn from the SLM (the one in italic).
(b) Role-playing: We prompt the same SLM to play two different roles (Appendix~\ref{subsubsection: Prompting an SLM to Form a  Dialogue}) and concatenate the speeches in each turn to form the audio of a dialogue and crop one minute for evaluation.
Note that the illustration above is not exactly how we prompt the SLM to play two roles.
Refer to Appendix~\ref{subsubsection: Prompting an SLM to Form a  Dialogue} for more details.}
\label{fig:dataset_illustration.pdf}
\end{figure*}

\section{Dataset Construction}

In this section, we introduce how we construct the StyleSet.
The dataset will be released based on the MIT license.

\subsection{Voice Style Instruction Following}
\label{appendix: subsection: dataset construction: Voice Style Instruction Following}
We formulate this task as a multi-turn dialogue between the user and the SLM.
The user first tells the SLM to speak a specific sentence without specifying the style, and the SLM should repeat the sentence.
Next, the user asks the SLM to speak the same sentence in a specific style.
We synthesize the user turn speeches with GPT-4o.
We split the semantic instruction and style instruction into two turns since cramming them into a single turn makes it hard for the SLM to understand what to do, based on preliminary experiments.

\subsection{Role-Playing}
\label{appendix: subsection: dataset construction: Role-Playing}

We construct the role-playing contexts from IEMOCAP~\citep{iemocap}, a dataset that contains scripted and improvised dyadic dialogues between two individuals.
The role-playing contexts we create include the background information of the dialogue, the relationship between the two speakers, and the transcription of the first turn.
We contact the authors of IEMOCAP and obtain their consent for redistributing the role-playing contexts in StyleSet.
We give the transcription of the first turn of the dialogue to make sure that the dialogues generated by different SLMs all start from the same line.

\subsubsection{Prompting an SLM to Form a  Dialogue}
\label{subsubsection: Prompting an SLM to Form a  Dialogue}
To evaluate an SLM, we prompt the SLM to play two different roles.
We call the SLM that plays the role 1 "SLM\textsubscript{1}" and the same SLM that plays the other role "SLM\textsubscript{2}".
Here, SLM\textsubscript{1} and SLM\textsubscript{2} are powered by the same SLM to be evaluated, but they are prompted to start the dialogue in a different way.
Precisely, SLM\textsubscript{1}, which plays the first role and starts with the specified line, is given the following input (in speech):

\texttt{Imagine the following situation: [dialoge\_context].}
\texttt{Now, pretend that you are [role\_1] and I am [role\_2]. Let's start the dialogue with you. Please start the dialogue with the following sentence and do not say anything else.}
\texttt{[first\_line]}

Given this prompt, SLM\textsubscript{1} should speak the first line.
For SLM\textsubscript{2}, we prompt it to engage in the role-playing dialogue with the following prompt (in speech):

\texttt{Imagine the following situation: [dialoge\_context].}
\texttt{Now, pretend that you are [role\_2] and I am [role\_2]. Engage in a dialogue with me while acting in the specified role.  I will start first.}
\texttt{[first\_line]}

Given the above input, SLM\textsubscript{2} should respond to the first line, and the dialogue can continue by using the output from SLM\textsubscript{2} as the input to SLM\textsubscript{1} and then using the out from SLM\textsubscript{1} as the input to SLM\textsubscript{2}.
The speech prompts used to initiate the role-playing dialogues are synthesized with 4o-audio.
We let SLM\textsubscript{1} and SLM\textsubscript{2} engage in a twenty-turn dialogue.

\subsubsection{Constructing the Audios for Evaluation}
We evaluate the dialogue created with an SLM by concatenating the audio of each turn separated by a two-second silence,\footnote{Currently, the dialogue created by the SLM turn-by-turn, i.e., $SLM_2$ speaks after $SLM_1$ has finished and vice versa. This turn-taking dialogue may be different from the full-duplex dialogue between humans. We do not consider full-duplex settings since most SLMs are not capable of doing this. However, the evaluation pipeline we propose naturally supports evaluating full-duplex dialogue.} and ask an ALLM judge to evaluate the dialogue.
While the silence between two turns is longer than the inter-turn silence in real-world dialogue, we deliberately use a longer silence since some SLMs we use only support a single voice, i.e., the voice of the two roles will be identical.
If we use a shorter silence interval to concatenate turns, the whole audio will sound like a monologue, making it hard to evaluate for the judges.
We crop the dialogue to a maximum of one minute (corresponding to at least three turns) for evaluation.

\subsection{Examples of the Dataset}
\label{subsection: Examples of the Dataset}

We prepare some examples of the dataset in Table~\ref{tab:voice-instruction} and Table~\ref{tab:roleplaying-contexts}.

\section{Evaluation Prompts and Instructions}
\label{appendix section: Evaluation Prompts and Instructions}

\subsection{Prompts for ALLM Judges}
\label{appendix section: Prompts for ALLM Judges}

The full prompt when using ALLM-as-a-judge are specified in Table~\ref{tab:tts_evaluation_prompt} for evaluating the voice style IF, Table~\ref{tab:role_play_eval_prompt} for evaluating the style in role-playing, and Table~\ref{tab:human_likeness_judgment} for evaluating the realism for role-playing.
Importantly, the evaluation prompts are all given in texts, i.e., the only audio input to the ALLM judge is the audio to be evaluated.
This is different from ~\citet{s2sarena}, which converts the evaluation prompts into speeches, and feeds the evaluation instruction speeches and the speech to be evaluated together into the ALLM judges.

\subsection{Details of Human Evaluation}
\label{subsection: Details of Human Evaluation}

\begin{figure*}[t]

\centering
\includegraphics[clip, trim = 0px 0px 0px 0px,width=0.9\linewidth]{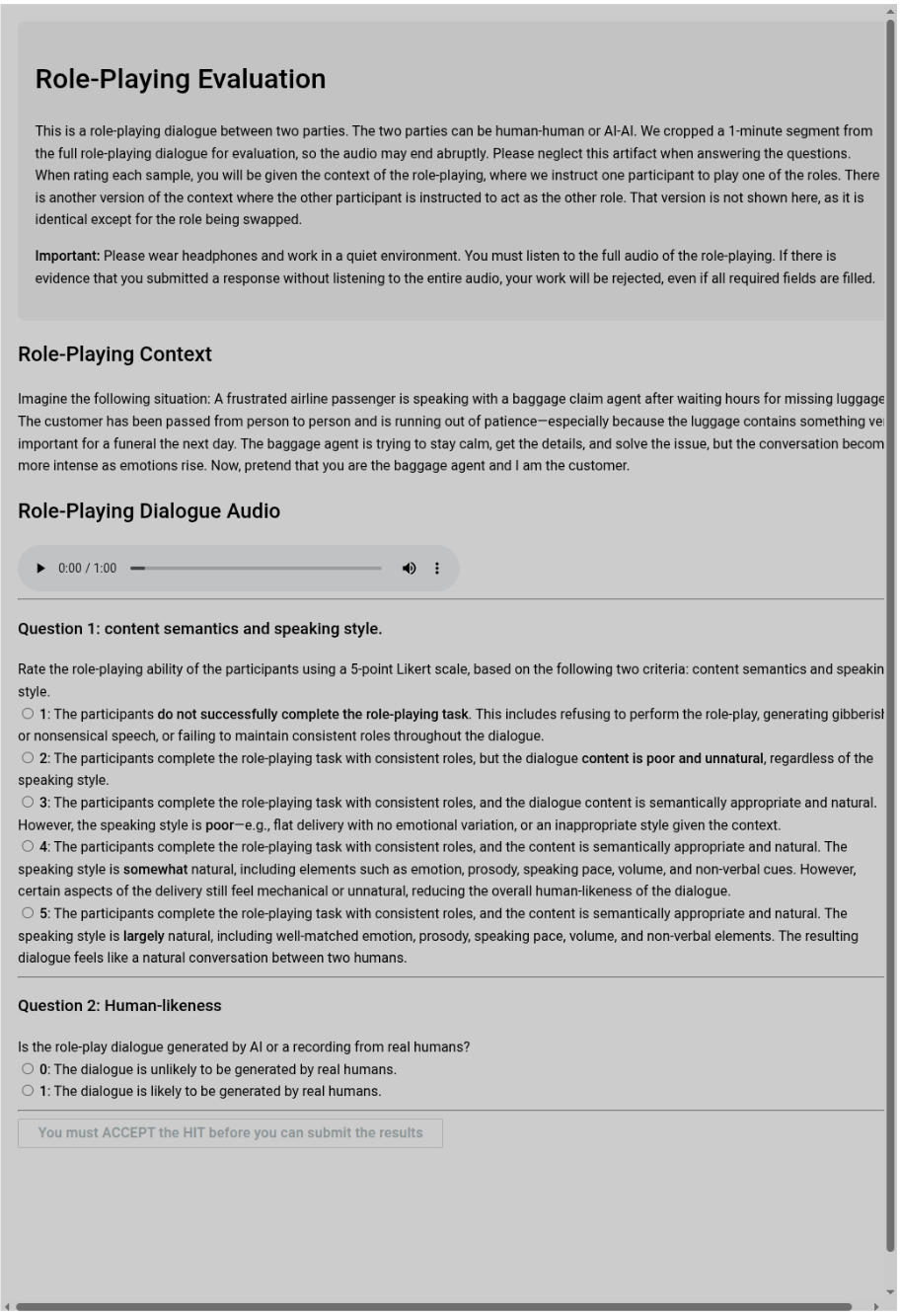}
\caption{The interface used to collect human evaluation results for the role-playing task.}
\label{fig:interface.png}
\end{figure*}

We use Prolific and an internal platform to recruit evaluators who are all native English speakers.
Each evaluator is paid an hourly wage of 12 USD to 24 USD for completing the task.
The evaluators are instructed that the tasks they are working on are related to academic research and their responses will be collected for research purposes.
They are also aware that participating in the task implies they agree to share the results with the research community. 
The instructions given to the human evaluators are mostly the same as the prompts shown to the ALLM judges, except we add some formatting instructions to the ALLM judges that are not given to the human evaluators.
An example of the interface used in the role-playing task is shown in Figure~\ref{fig:interface.png}.
Note that the detailed task instructions are given to the human evaluators using a separate Google document; we omit those instructions here since they are mostly the same as those given to the ALLM judges.

\section{Experiment Details}
\label{section: Experiment Details}

When generating the responses from SLMs, we use the following hyperparameters: 
We set the temperature to 1.0 and the nucleus sampling threshold (\texttt{top\_p}) to 0.9~\citep{neucleus}. 
The maximum number of newly generated tokens (\texttt{max\_new\_tokens}) is 4096. 
We fix the random seed to 42 for reproducibility.

When generating the evaluation responses from ALLM judges, we use the hyperparameters: 
We set the decoding temperature to 1.0 and use nucleus sampling with a threshold of $\texttt{top\_p} = 0.9$. 
The maximum number of newly generated tokens is limited to \texttt{256} via the \texttt{max\_new\_tokens} parameter.

The 4o-audio and 4o-mini audio models we use are the 2024-12-17 version.

\begin{table*}[h]
    \footnotesize
    \centering
    \begin{tabular}{|p{50em}|}
        \hline
\texttt{\# Task Introduction}\\
\texttt{}\\
\texttt{You are a helpful assistant. Your task is to evaluate the quality of speech generated by a text-to-speech (TTS) model.}\\
\texttt{The model is given a specific piece of text and a target speaking style; the model is asked to repeat the text in the target speaking style.}\\
\texttt{Your job is to assess whether the generated speech accurately follows the given text and expresses the intended style.}\\
\texttt{}\\
\texttt{\# Evaluation Criteria}\\
\texttt{}\\
\texttt{Rate the generated speech using a 5-point Likert scale based on two criteria:}\\
\texttt{}\\
\texttt{1. **Text Following**:}\\
\texttt{   Does the generated speech accurately follow the text it was instructed to say?}\\
\texttt{   - If the speech omits required words or includes additional content not present in the original text, it should be considered as **not following** the text.}\\
\texttt{}\\
\texttt{2. **Style Following**:}\\
\texttt{   Does the generated speech follow the style specified in the instruction?}\\
\texttt{   - Style may include emotional tone, prosody, emphasis, or any specific expressive requirement.}\\
\texttt{}\\
\texttt{\#\#\# Scoring Rubric}\\
\texttt{}\\
\texttt{- **1**: The speech **does not follow** the required text, regardless of style.}\\
\texttt{- **2**: The speech **follows the text** but **does not follow any** part of the style instruction.}\\
\texttt{- **3**: The speech **follows the text**, but less than half of the style instructions are **perfectly satisfied**.}\\
\texttt{- **4**: The speech **follows the text**, and more than half of the style instructions, but not all of the style instructions, are **perfectly satisfied**.}\\
\texttt{- **5**: The speech **follows the text** and **fully follows** the style instruction.}\\
\texttt{}\\
\texttt{\# Evaluation Steps}\\
\texttt{}\\
\texttt{1. **Analyze the Instruction**}\\
\texttt{   Carefully read the style instruction and list all style-related requirements that the TTS-generated speech must satisfy.}\\
\texttt{}\\
\texttt{2. **Check Text Accuracy**}\\
\texttt{   Listen to the generated speech and check whether the spoken text exactly matches the given text.}\\
\texttt{   - If it does **not** match, assign a score of **1** and skip the style evaluation.}\\
\texttt{}\\
\texttt{3. **Evaluate Style Adherence**}\\
\texttt{   For each requirement you listed in Step 1, evaluate how well the speech satisfies it. Provide a brief analysis of how the speech does or does not meet each requirement.}\\
\texttt{}\\
\texttt{4. **Assign a Style Score**}\\
\texttt{   Based on your analysis, assign a final score between **2 and 5** reflecting how well the speech follows the intended style. Follow the scoring rubric strictly.}\\
\texttt{}\\
\texttt{5. **Report the Final Score**}\\
\texttt{   Conclude your evaluation with the following format:  \texttt{Final score: [[score]]}. Replace \texttt{score} with an integer in \{1, 2, 3, 4, 5\}. Keep the brackets as shown.}\\
\texttt{}\\
\texttt{\# Text for the TTS Model to Generate Speech}\\
\texttt{}\\
\texttt{\{target\_text\}}\\
\texttt{}\\
\texttt{\# Speaking Style Instructions to the TTS Model}\\
\texttt{}\\
\texttt{\{speaking\_style\}}\\
\texttt{}\\
\texttt{\# The Speech Generated by the TTS Model}\\
        \hline
    \end{tabular}
    \caption{The evaluation prompt for voice style IF.
    While the prompt says we are evaluating the TTS model, we are actually evaluating the outputs of SLMs.
    Nevertheless, for the judge model, it does not matter whether the speeches are generated by a TTS model or an SLM conducting the voice style IF task.}
    \label{tab:tts_evaluation_prompt}
\end{table*}

\begin{table*}[h]
    \tiny
    \centering
    \begin{tabular}{|p{75em}|}
        \hline
\texttt{\# Task Introduction}\\
\texttt{}\\
\texttt{You are a helpful assistant. Your task is to evaluate the role-playing ability of a voice assistant. The voice assistant is given a context background that includes two roles, and it is asked to act as both roles and generate a spoken dialogue. You need to assess the quality of the voice assistant’s role-playing performance.}\\
\texttt{}\\
\texttt{\# Evaluation Criteria}\\
\texttt{}\\
\texttt{Rate the role-playing ability of the voice assistant using a 5-point Likert scale, based on the following two criteria:}\\
\texttt{}\\
\texttt{1. **Content Semantics**}\\
\texttt{   This criterion evaluates the semantic quality of the spoken content. Consider whether the dialogue is aligned with the background context, whether the conversation includes repetitive or irrelevant information, and whether the wording is natural given the relationship between the two roles and the topic being discussed.}\\
\texttt{}\\
\texttt{2. **Speaking Style**}\\
\texttt{   This criterion evaluates the non-textual aspects of the spoken dialogue. Consider whether the emotion and speaking style are appropriate for the context, and whether non-verbal elements (e.g., sighs, laughter, short pauses) are used effectively to enhance naturalness. Also assess whether the prosody, intonation, speaking pace, and volume sound natural and appropriate.}\\
\texttt{}\\
\texttt{\#\#\# Scoring Rubric}\\
\texttt{}\\
\texttt{- **1**: The voice assistant does not successfully complete the role-playing task. This includes refusing to perform the role-play, generating gibberish or nonsensical speech, or failing to maintain consistent roles throughout the dialogue.}\\
\texttt{}\\
\texttt{- **2**: The voice assistant completes the role-playing task with consistent roles, but the dialogue content is poor and unnatural, regardless of the speaking style.}\\
\texttt{}\\
\texttt{- **3**: The voice assistant completes the role-playing task with consistent roles, and the dialogue content is semantically appropriate and natural. However, the speaking style is poor—e.g., flat delivery with no emotional variation, or an inappropriate style given the context.}\\
\texttt{}\\
\texttt{- **4**: The voice assistant completes the role-playing task with consistent roles, and the content is semantically appropriate and natural. The speaking style is **somewhat** natural, including elements such as emotion, prosody, speaking pace, volume, and non-verbal cues. However, certain aspects of the delivery still feel mechanical or unnatural, reducing the overall human-likeness of the dialogue.}\\
\texttt{}\\
\texttt{- **5**: The voice assistant completes the role-playing task with consistent roles, and the content is semantically appropriate and natural. The speaking style is **largely** natural, including well-matched emotion, prosody, speaking pace, volume, and non-verbal elements. The resulting dialogue feels like a natural conversation between two humans.}\\
\texttt{}\\
\texttt{\# Evaluation Steps}\\
\texttt{}\\
\texttt{Your response should include the following four steps:}\\
\texttt{}\\
\texttt{1. **Analyze the Role and Context**}\\
\texttt{   Carefully read the context and understand the two roles the voice assistant is expected to play.}\\
\texttt{}\\
\texttt{2. **Listen to the Role-Play Dialogue**}\\
\texttt{   Listen closely to the spoken dialogue generated by the voice assistant.}\\
\texttt{}\\
\texttt{3. **Analyze the Role-Play Dialogue Based on the Evaluation Criteria**}\\
\texttt{   Evaluate the quality of the role-play according to the criteria above. Reference specific turns or moments in the dialogue to support and justify your evaluation.}\\
\texttt{}\\
\texttt{4. **Report the Final Score**}\\
\texttt{   Conclude your evaluation with the following format:}\\
\texttt{   \texttt{Final score: [[score]]}}\\
\texttt{   Replace \texttt{score} with an integer in \{1, 2, 3, 4, 5\}. Keep the double brackets as shown.}\\
\texttt{}\\
\texttt{\# Context of the Role-Playing}\\
\texttt{}\\
\texttt{\{role\_play\_context\}}\\
\texttt{}\\
\texttt{**Note**: In the above context, the voice assistant is instructed to act as one of the roles. There is another version of the context where the assistant is instructed to act as the other role. That version is not shown here, as it is identical except for the role being swapped.}\\
\texttt{}\\
\texttt{\# The Spoken Dialogue of the Role-Playing}\\
        \hline
    \end{tabular}
    \caption{The evaluation prompt for the \textit{style} aspect of the role-playing task.}
    \label{tab:role_play_eval_prompt}
\end{table*}

\begin{table*}[h]
    \footnotesize
    \centering
    \begin{tabular}{|p{50em}|}
        \hline
\texttt{\# Task Introduction}\\
\texttt{}\\
\texttt{You are a helpful assistant. Your task is to judge if a spoken dialogue is generated by AI or a recording from real humans. The spoken dialogue is a role playing.}\\
\texttt{}\\
\texttt{\# Evaluation Criteria}\\
\texttt{}\\
\texttt{Rate the dialogue using a binary 0/1 decision.}\\
\texttt{}\\
\texttt{- **0**: The dialogue is unlikely to be generated by real humans.}\\
\texttt{}\\
\texttt{- **1**: The dialogue is likely to be generated by real humans.}\\
\texttt{}\\
\texttt{\# Evaluation Steps}\\
\texttt{}\\
\texttt{Your response should include the following four steps:}\\
\texttt{}\\
\texttt{1. **Analyze the Role and Context**}\\
\texttt{   Carefully read the context and understand the two roles in the role-playing.}\\
\texttt{}\\
\texttt{2. **Listen to the Role-Play Dialogue**}\\
\texttt{   Listen closely to the spoken dialogue.}\\
\texttt{}\\
\texttt{3. **Analyze the Role-Play Dialogue**}\\
\texttt{   Evaluate the role-play according to whether it is likely generated by humans.}\\
\texttt{}\\
\texttt{4. **Report the Final Score**}\\
\texttt{   Conclude your evaluation with the following format:}\\
\texttt{   \texttt{Final score: [[score]]}}\\
\texttt{   Replace \texttt{score} with an integer in \{0, 1\}. Keep the double brackets as shown.}\\
\texttt{}\\
\texttt{\# Context of the Role-Playing}\\
\texttt{}\\
\texttt{\{role\_play\_context\}}\\
\texttt{}\\
\texttt{**Note**: In the above context, the participant is instructed to act as one of the roles. There is another version of the context where the other participant is instructed to act as the other role. That version is not shown here, as it is identical except for the role being swapped.}\\
\texttt{}\\
\texttt{\# The Spoken Dialogue of the Role-Playing}\\
        \hline
    \end{tabular}
    \caption{Evaluation prompt for the \textit{realism} aspect for role-playing.}
    \label{tab:human_likeness_judgment}
\end{table*}

\begin{table*}[ht]
    \centering
    \footnotesize
    \begin{adjustbox}{max width=\textwidth}
        \begin{tabular}{p{8cm} p{7cm}}
            \toprule
            \textbf{Text to speak} & \textbf{Speaking style} \\
            \midrule
            I don’t know! I swear I didn’t see anything last night. Let me go, please! & Say it in a trembling voice and a frightened emotion. \\
            \midrule
            I can’t believe this. I still remember the first day she came here, but now she is gone. & Speak in a sobbing voice. Start with moderate volume and gradually fade until the last word is almost unhearable. \\
            \midrule
            I was just, uh, just out with Jake, and we were gonna come back earlier, but his car wouldn’t start and then—then my phone died, so I couldn’t call you… & Speak with a nervous and frightened tone, including natural stuttering. \\
            \midrule
            Oh, great. Because that’s exactly what we needed right now—another disaster. & Speak in a sarcastic tone. Prolong “Oh”, emphasize “exactly”, and insert a pause before “another disaster”. Speak at a moderately fast pace. \\
            \midrule
            Oh my god! Oh my god! Oh my god! I can’t believe this is true! & Speak in an ecstatic tone. Start with a 2-second scream, say the first “Oh my god” slowly, the second louder and faster, and the third nearly shouted. Say the last part fast and excitedly, ending with laughter. \\
            \midrule
            Who’s a happy baby? You are! Yes, you are! Look at those little toes! Teeny tiny toes! & Speak using motherese — speak in a high pitch, with exaggerated intonation and slower tempo. \\
            \bottomrule
        \end{tabular}
    \end{adjustbox}
    \caption{Voice style IF examples.}
    \label{tab:voice-instruction}
\end{table*}

\begin{table}[ht]
    \centering
    \footnotesize
    \begin{adjustbox}{max width=\textwidth}
        \begin{tabular}{p{16cm}}
            \toprule
            \textbf{Dialogue context} \\
            \midrule
            Imagine the following situation: There is a couple with a newborn baby, Amy. One day, the wife is called to enroll in the army in a foreign country. She has to separate from her spouse for more than 1 year. The couple are sad about this. Now, pretend that you are the husband and I am the leaving wife. Engage in a dialogue with me while acting in the specified role. Let's start the dialogue from you. Please start the dialogue with the following sentence and do not say anything else. So you're leaving tomorrow. \\
            \midrule
            Imagine the following situation: A customer and a customer service are talking. The customer has been talking to a machine and he is finally transferred to an operator. The customer is frustrated, and the operator is trying to change the mood of the customer and solve the problem. Now, pretend that you are the operator and I am the customer. Engage in a dialogue with me while acting in the specified role. Let's start the dialogue from you. Please start the dialogue with the following sentence and do not say anything else. Hello. This is Viacom Services. How can I help you? \\
            \midrule
            Imagine this situation: A son and his mother are talking in the backyard early in the morning. His older brother, Larry, went missing in the war three years ago. The father still believes Larry is alive, but the son thinks it's time to accept that he's gone. He wants to move on with his life and marry Larry’s old girlfriend, Annie. The mother doesn’t want to let go of hope, and she’s upset. Now, pretend that I am the son and you are the mother. Engage in a dialogue with me while acting in the specified role. Let's start the dialogue from you. Please start the dialogue with the following sentence and do not say anything else. What's he going to say? Maybe we should tell him before he sees it. \\
            \midrule
            Imagine the following situation: There is a couple with a newborn baby, Amy. One day, the wife is called to enroll in the army in a foreign country. She has to separate from her spouse for more than 1 year. The couple are sad about this. Now, pretend that I am the husband and you are the leaving wife. Engage in a dialogue with me while acting in the specified role. Let's start the dialogue from you. Please start the dialogue with the following sentence and do not say anything else. How's it going, babe? Yeah? \\
            \midrule
            Imagine the following situation: A frustrated airline passenger is speaking with a baggage claim agent after waiting hours for missing luggage. The customer has been passed from person to person and is running out of patience—especially because the luggage contains something very important for a funeral the next day. The baggage agent is trying to stay calm, get the details, and solve the issue, but the conversation becomes more intense as emotions rise. Now, pretend that you are the baggage agent and I am the customer. Engage in a dialogue with me while acting in the specified role. Let's start the dialogue from you. Please start the dialogue with the following sentence and do not say anything else. Can I help you sir? \\
            \bottomrule
        \end{tabular}
    \end{adjustbox}
    \caption{Role-playing contexts for role-playing tasks.}
    \label{tab:roleplaying-contexts}
\end{table}

\end{document}